\documentclass[epj,nopacs]{svjour}
\usepackage{amsmath, amsfonts, amssymb, eufrak}
\usepackage{latexsym}
\usepackage{graphics}

\newcommand{\ds}{\displaystyle}
\newcommand{\const}{{\rm const}}
\renewcommand{\H}{{\cal H\,}}

\newcommand{\A}{\mathcal{A}}

\newcommand{\M}{\mathbb{M}}
\newcommand{\1}{\mathbb{I}}

\newcommand{\set}[2]{\{ #1:\,#2 \}}

\newcommand{\SM}{\mbox{$^{s}\mathbb{M}$}}
\newcommand{\iso}[2]{{\sf ISO}(#1, #2)}
\newcommand{\siso}[2]{\mbox{$^{s}{\sf ISO}(#1, #2)$}}
\newcommand{\isoa}[2]{{\mathfrak{iso}}(#1, #2)}
\newcommand{\so}[2]{{\sf SO}(#1, #2)}
\newcommand{\T}[1]{{\sf T}_{#1}}

\begin{document}
%
%
\title{Quantum mechanics on Riemannian Manifold in Schwinger's
       Quantization Approach IV}

\subtitle{Quantum mechanics of Superparticle}
\author{Chepilko Nicolai Mikhailovich\inst{1}
\and Romanenko Alexander Victorovich\inst{2}} \institute{Physics
Institute of the Ukrainian Academy of Sciences, Kyiv-03 028,
Ukraine, \email{chepilko@zeos.net}
\and Kyiv Taras Shevchenko
University, Department of Physics, Kyiv-03 022, Ukraine,
\email{ar@ups.kiev.ua}}

\authorrunning{N. M. Chepilko, A. V. Romanenko}

\titlerunning{Quantum mechanics on Riemannian Manifold, IV}
\date{Received: date / Revised version: date}

\abstract{%
In this paper we extend Schwinger's quantization approach to the case of
a supermanifold considered as a coset space of the Poincare group by the
Lorentz group. In terms of coordinates parametrizing a supermanifold,
quantum mechanics for a superparticle is constructed. As in models
related to the usual Riemannian manifold,
the key role in analyzes is played by Killing vectors. The main feature of
quantum theory on the supermanifold consists of the fact that the spatial
coordinates are not commute with each other and therefore  are represented
on wave functions by integral operators.
\PACS{
      {PACS-key}{discribing text of that key}   \and
      {PACS-key}{discribing text of that key}
     } 
} 
\maketitle
\section{Introduction}

The present paper accomplishes our series of works
(see \cite{paper1}--\cite{paper3}) devoted to extending Schwinger's
quantization procedure to the case of quantum theory
defined on manifolds with a group structure. The aim of the present paper
is to develop quantum mechanics for a particle moving on a superspace
treated as a coset space
$$
\SM^{n}\simeq\siso{p}{q}/\so{p}{q},\: \mbox{where}\: p+q=n\,,
$$
where $\so{p}{q}$ denotes a special orthogonal rotation group and  $\siso{p}{q}$
is a supersymmetrical extension of the Poin\-ca\-re group $\iso{p}{q}$.
A local coordinate system in a superspace consists of the variables
$z^{A}=(x^{\mu}, \psi, \overline{\psi})$, where $x^{\mu}$, $\mu=\overline{1, n}$
denote spatial coordinates of underlying pseudoeuclidean space
$\M^{n}\simeq\iso{p}{q}/\so{p}{q}$, and $\psi$,~${\overline{\psi}}$
are Dirac bispinors, whose inner structure is in accordance with the dimension
and the signature of underlying manifold $\M^{n}$. We have briefly examined
the main features of the definition of a superspace in section \ref{s1}.

The quantum mechanics of a superparticle is construc\-ted in terms of
a logical scheme presented in \cite{paper1}--\cite{paper3}.
To obtain the explicit form of permissible variations of coordinates we have
analyzed the symmetries of the Lagrangian which determines the dynamical
behavior of a superparticle. The results obtained at the classical level
are briefly discussed in sections \ref{s2} and \ref{s3}.
As it has appeared in \cite{paper1}--\cite{paper3}, the permissible variations
$\delta z^{A}$ are closely related to Killing vector fields defined, however,
on $\SM^{n}$ instead of $\M^{n}$.

In section \ref{s4} the generator of permissible variations is constructed.
Using it we derive the commutation relations for quantum-mechanical
operators that are presented in section \ref{s5}. The main feature of the
algebra of commutation relations is that the coordinates $z^{A}$
(including the spatial ones) do not commute with each other. The fact that
$[x^{\mu}, x^{\nu}]\ne 0$ means thatthe space-time coordinates $x^\mu$
in quantum theory lose the meaning which they have in $\M^4$.
The presence of odd variables $\psi$ and $\overline{\psi}$
modifies the geometric nature of space-time coordinates.
In section \ref{s5} we also discuss the question about constructing the
total set of commuting variables which determines the way of description
of the physical space of states.

Using the total set given in section \ref{s5}, we develop the
coordinate representation in section \ref{s6} .  Namely, we obtain the
representation of operators on the space of wave functions and the form
of Schr\"odinger equation. The essential feature of the coordinate
representation consists of the fact that the spatial coordinates
$x^{\mu}$ are represented as integral operators. Possibly, this result
points out to existence of some effective spatial size of a
superparticle. This fact should explain the decrease of divergences in
integrals of motion in supersymmetrical quantum field theories.


\section{Supersymmetry and Superspace}
\label{s1}

Supersymmetry realizes the formal connection between bosonic and
fermionic states of quantum systems. The basic principles of a
supersymmetry have been discovered in 1970s in the series of works
\cite{volkov}, \cite{wess}.

The aim of the present section is to examine the definition of a
superspace and explain some notation used in text. As in the case of
other symmetries, the sypersymmetry can be formulated in terms of a
transformation group acting on physical states. But, taking into
account the features of fermi statistics, one can observe that the
framework of classical Lie groups and algebras become inconsistent
in description of a supersymmetry. Such a problem can be solved with
the use of $\mathbb{Z}_{2}$-graded Lie groups and algebras (see
\cite{berezin}).

$\mathbb{Z}_{2}$-grading of the algebra $\A$ means the existence of a
map
$$
\mathfrak{a}:\A\longrightarrow\mathbb{Z}_{2}\,,
$$
where $\mathbb{Z}_{2}\equiv \{0,1\}$.
The algebra $\A$ can be divided into two parts: the {\it even} elements
$\A_{e}=\set{a\in\A}{\mathfrak{a}(a)=0}$ and the {\it odd} elements
$\A_{o}=\set{a\in\A}{\mathfrak{a}(a)=1}$, so that $\A=\A_{e}\oplus\A_{o}$.

The definition of a Lie bracket can be modified for the case of
graded algebras as in following
\begin{equation}
[a, b]:=ab-s(a, b)ba\,,\qquad\forall a, b\in\A
\label{1-1}
\end{equation}
where $s(a, b)$ denotes the signature factor
$$
s(a, b)=(-1)^{\mathfrak{a}(a)\cdot\mathfrak{a}(b)}=
\left\{
\begin{array}{rl}
-1\,,& \mbox{when $a$, $b$ are odd}\\
+1\,,& \mbox{in all the other cases}
\end{array}
\right.
$$
The generalized Lie bracket is called the {\it supercommutator}
and satisfies the following identities
\begin{align}
[a, b]&=-s(a, b)[b, a]\,,\nonumber \\
[a, b_{1}+b_{2}]&= [a, b_{1}]+[a, b_{2}]\,,\nonumber \\
[a, bc]&=[a, b]c+s(a, b) b [a, c]\,,\\
s(a, c)[a, [b, c]]&+ s(b, a)[b, [c, a]]+s(c, b)[c, [a, b]]=0\,.\nonumber
\label{1-2}
\end{align}

In the flat space the physical features of a theory are determined by
transformation properties of the objects under the
actions of the group of symmetries of the pseudoeuclidean space
$\M^{n}$ equipped with the diagonal metric
$$ \{\eta_{\mu\nu}\}={\rm diag}(\underbrace{1,\dots, 1}_{p}, \underbrace{-1,\dots,
-1}_{q})\,,\quad \mu, \nu=\overline{1, n}\,.
$$
This group is $\ds\frac{n(n+1)}{2}$-parameter inhomogeneous
Lie group $\iso{p}{q}$. In the case
$n=4$, $p=1$, $q=3$ and
$$
\eta={\rm diag}(1, -1, -1, -1)
$$
it coincides with $\iso{1}{3}=\so{1}{3}\ltimes {\sf T}_{4}$,
and the corresponding Lie algebra satisfies the following commutation
relations
\begin{subequations}
\label{1-3}
\begin{align}
[P_{\mu}, P_{\nu}]&=0\,,\\
[J_{\mu\nu}, P_{\alpha}]&=i(\eta_{\mu\alpha}P_{\nu}-\eta_{\nu\alpha}P_{\mu})\,,\\
[J_{\mu\nu}, J_{\alpha\beta}]&=i\left(
\eta_{\mu\alpha}J_{\nu\beta}+\eta_{\nu\beta}J_{\mu\alpha}-\eta_{\nu\alpha}J_{\mu\beta}-\eta_{\mu\beta}J_{\nu\alpha} \right)\,.
\end{align}
\end{subequations}
Here $P_{\mu}$ and $J_{\mu\nu}$ denote the generators of translations
and $\so{1}{3}$-rotations correspondingly. Due to the structure of
semidirect product of the Poincare group, the space $\M^{4}$ is
isomorphic to ${\sf T}_{4}$ and therefore can be defined as a
quotient space
\begin{equation}
\M^{4}\simeq
\iso{1}{3}/\so{1}{3}
\label{1-4}
\end{equation}
Denote the coordinates of $\M^{4}$ by $\{x^{\mu}\}$. Then the transformations
generated by
$$
\varepsilon^{\mu}P_{\mu}+\frac{1}{2}\varepsilon^{\mu\nu}J_{\mu\nu}\in\isoa{1}{3}\,,
$$
with
$\varepsilon^{\mu\nu}=-\varepsilon^{\nu\mu}=\const$, $\varepsilon^{\mu}=\const$,
causes the following changes of space-time coordinates
\begin{equation}
x^{\mu}\to {x'}^{\mu}=x^{\mu}+\varepsilon^{\mu}+\varepsilon^{\mu}{}_{\nu}x^{\nu}\,.
\label{1-5}
\end{equation}

The main idea on which the supersymmetry is based consists of the usage
in (\ref{1-4}) the extension of the group $\iso{1}{3}$ instead of $\iso{1}{3}$.
Such an extension can be developed by adding to the Lie algebra
new generators that must correspond to some representation of
$\so{1}{3}$ . The simplest way of doing it is to include spinor generators
$Q$, $\overline{Q}$ with anticommutative components with the following
properties
\begin{subequations}
\label{1-6}
\begin{align}
[P_{\mu}, Q_{a}]&=0\,,\quad [P_{\mu}, \overline{Q}^{a}]=0\,,\\
[Q_{a}, J_{\mu\nu}]&=-(\sigma_{\mu\nu})^{a}{}_{b}Q^{b}\,,\quad
[\overline{Q}^{a}, J_{\mu\nu}]=\overline{Q}^{b}(\sigma_{\mu\nu})^{a}{}_{b}\,,\quad\\
[Q_{a}, \overline{Q}^{b}]&=(\gamma^{\mu})_{a}{}^{b}P_{\mu}\,.
\end{align}
\end{subequations}
where $\sigma_{\mu\nu}=\frac{i}{4}[\gamma_{\mu},\gamma_{\nu}]$.
The algebra (\ref{1-3}), (\ref{1-6}) coincides with $N=1$ supersymmetrical
extension of the Poincare algebra. The coset space
$$
\SM^{4}\simeq\siso{1}{3}/\so{1}{3}\,.
$$
is called a {\it superspace} (here $\siso{1}{3}$ denotes the
corresponding extended group). The coordinates of $\SM^{4}$ are denoted
as $(x^{\mu}, \psi, \overline{\psi})$, where $\psi$ and
$\overline{\psi}\equiv\psi^{\dagger}\gamma^{o}$ are Dirac bispinors
with anticommutative components, $x^{\mu}$ are the coordinates of the
space $\M^{4}$, that is called an underlying space for $\SM^{4}$.

After introducing the unified notation $z^{A}=(x^{\mu}, \psi, \overline{\psi})$
and taking into account the anticommutativity of
$\psi$ and $\overline{\psi}$ we can write
\begin{equation}
z^{A}\cdot z^{B}=s(z^{A}, z^{B})z^{B}\cdot z^{A}\,.
\label{1-7}
\end{equation}

The transformations induced by the generator
$G=\overline{\xi}Q+\overline{Q}\xi$, where $\xi$,
$\overline{\xi}$ are Grassmanian spinor constants, have the form
\begin{subequations}
\label{1-9}
\begin{align}
x^{\mu}&\longrightarrow {x'}^{\mu}=x^{\mu}+\frac{i}{2}\left( \overline{\psi}\gamma^{\mu}\xi-\overline{\xi}\gamma^{\mu}\psi \right)\,,\\
\psi&\longrightarrow \psi'=\psi+\xi\,,\quad  \overline{\psi}\longrightarrow \overline{\psi}\,'=\overline{\psi}+\overline{\xi}\,.
\end{align}
\end{subequations}
These transformations are called  {\it supertranslations}.
One can observe from (\ref{1-6}) that supertranslations commute with the
usual translations generated by $\varepsilon^{\mu}P_{\mu}$.

In the superspace with the coordinates $z^{A}=(x^{\mu}, \psi,
\overline{\psi})$ 1-forms $dx^{\mu}$ are not invariant under
supertranslations (\ref{1-9}) and therefore cannot be chosen as a
basis of tensor fields. Let us consider the transformation laws of
$dz^{A}$  in the case of supertranslations with the parameters $\xi$ and
$\overline{\xi}$. Since
$$ dz^{A}\longrightarrow
{dz'}^{A}=\frac{\partial {z'}^{A}}{\partial z^{B}}dz^{B}\,,
$$
then
$$ d\psi\longrightarrow d\psi\,,\quad
d\overline{\psi}\longrightarrow d\overline{\psi}\,,
$$
$$
dx^{\mu}\longrightarrow dx^{\mu}+\frac{i}{2}\left(
d\overline{\psi}\gamma^{\mu}\xi-\overline{\xi}\gamma^{\mu}d\psi
\right)\,.
$$
Using the trivial equalities
$\delta\psi=\psi'-\psi=\xi$,
$\delta\psi=\overline{\psi}\,'-\overline{\psi}=\overline{\xi}$ we
arrive to the conclusion that invariant under supertranslations
(\ref{1-9}) 1-forms coincide with
\begin{equation}
e^{\mu}:=dx^{\mu}+\frac{i}{2}\left( d\overline{\psi}\gamma^{\mu}\psi-\overline{\psi}\gamma^{\mu}d\psi \right)\,.
\label{1-10}
\end{equation}
Together with $d\psi$ and $d\overline{\psi}$ they form the basis of invariant
1-forms on $\SM^{4}$.

A similar analyzes can be carried out for the vector fields
$\dfrac{\partial }{\partial z^{A}}$. The transformation law reads
$$
\frac{\partial }{\partial z^{A}}=\frac{\partial z^{\prime B}}{\partial z^{A}}
\frac{\partial }{\partial z^{\prime B}}
$$
Hence
\begin{equation*}
\begin{gathered}
\frac{\partial }{\partial x^{\mu}}=\frac{\partial }{\partial x^{\prime\mu}}\,,\quad
\frac{\partial }{\partial \psi}=\frac{\partial }{\partial \psi'}-
\frac{i}{2}\overline{\xi}\gamma^{\mu}\frac{\partial }{\partial x^{\prime
\mu}}\,,\\
\frac{\partial }{\partial \overline{\psi}}=\frac{\partial }{\partial \overline{\psi}\,'}+
\frac{i}{2}\gamma^{\mu}\xi\frac{\partial }{\partial x^{\prime\mu}}\,.
\end{gathered}
\end{equation*}
Using
$\delta\psi=\psi'-\psi=\xi$, $\delta\overline{\psi}=\overline{\psi}\,'-\overline{\psi}=\overline{\xi}$
we find the following invariant vector fields
\begin{equation}
\overline{D}=\frac{\partial }{\partial \psi}-\frac{i}{2}\overline{\psi}\gamma^{\mu}\frac{\partial }{\partial x^{\mu}}\,,\quad
D=\frac{\partial }{\partial \overline{\psi}}+\frac{i}{2}\gamma^{\mu}\psi\frac{\partial }{\partial x^{\mu}}\,.
\label{1-11}
\end{equation}
Together with $\ds\frac{\partial }{\partial x^{\mu}}$  they form the basis of invariant
vector fields on $\SM^{4}$.

The expressions for the invariant 1-forms can be presented as
\begin{equation}
e^{A}=E^{A}_{B}dz^{B}\,,\quad e_{A}=\overline{E}^{B}_{A}\frac{\partial }{\partial z^{B}}\,,
\label{1-12}
\end{equation}
where
\begin{equation}
\begin{gathered}
\{E^{A}_{B}\}=
\left(
\begin{array}{ccc}
\delta^{\mu}_{\nu}&-\dfrac{i}{2}(\overline{\psi}\gamma^{\mu})^{a}&-\dfrac{i}{2}(\gamma^{\mu}\psi)_{a}\\[2mm]
0&\delta^{a}_{b}&0\\[2mm]
0&0&\delta^{a}_{b}
\end{array}
\right)\,,\quad\\[5mm]
\{\overline{E}^{A}_{B}\}=
\left(
\begin{array}{ccc}
\delta^{\mu}_{\nu}&0&0\\[2mm]
-\dfrac{i}{2}(\overline{\psi}\gamma^{\mu})^{a}&\delta^{a}_{b}&0\\[2mm]
\dfrac{i}{2}(\gamma^{\mu}\psi)_{a}&0&\delta^{a}_{b}
\end{array}
\right)\,.
\label{1-13}
\end{gathered}
\end{equation}
Here we assume that the differentials $d\psi$, $d\overline{\psi}$
are right multipliers in (\ref{1-12}).

These constructions can be easily generalized for the $n>4$
dimensional case. The superspace $\SM^{n}$ with the coordinates
$(x^{\mu}, \psi, \overline{\psi})$ is described by the coordinates of the
underlying manifold $\SM^{n}$ and $2^{[n/2]}$-component Dirac bispinors
with the Grassmanian components. The Clifford algebra is
generated by the relation for the basic elements
\begin{equation}
\gamma^{\mu}\gamma^{\nu}+\gamma^{\nu}\gamma^{\nu}=2\eta^{\mu\nu}\,,
\label{1-14}
\end{equation}
where $\eta_{\mu\nu}$ is the metric of $\M^{n}$.

The supertranslations have the form
\begin{align}
x^{\mu}\longrightarrow x^{\prime\mu}&+\frac{i}{2}\left( \overline{\psi}\gamma^{\mu}\xi-\overline{\xi}\gamma^{\mu}\psi \right)\,,\\
\psi\longrightarrow \psi'=\psi+\xi\,,&\quad \overline{\psi}\longrightarrow \overline{\psi}\,'=\overline{\psi}+\overline{\xi}\,.
\label{1-15}
\end{align}
In $n$-dimensional case the invariant 1-forms and vector fields
coincide with (\ref{1-10}) and (\ref{1-11}).


\section{Lagrangian for point particle in superspace}
\label{s2}

As an initial conception we choose the Maurer-Cartan 1-form (\ref{1-10}).
The velocity of the a superparticle can be defined in terms of this form as
following
\begin{equation*}
e^{\mu}=v^{\mu}d\tau\,,
\end{equation*}
where $\tau$  denotes the evolution parameter (for example, the physical time).
After introducing the symbol ``$\partial\,$'' that denotes the derivative
with respect to $\tau$, the velocity can be rewritten as
\begin{equation}
v^{\mu}=\partial x^{\mu}+\frac{i}{2}\left(\overline{\psi}\gamma^{\mu}\partial\psi-\partial\overline{\psi}\gamma^{\mu}\psi \right)\,,
\label{2-1}
\end{equation}
where the coordinates $(x^{\mu}, \psi, \overline{\psi})$  describing the
trajectory of a superparticle in a superspace are the functions of $\tau$.
Treating (\ref{2-1}) as a vector on the underlying manifold $\M^{n}$
let us define the Lagrangian as
\begin{equation}
L=\frac{1}{2}\eta_{\mu\nu}v^{\mu}v^{\nu}\,.
\label{2-2}
\end{equation}

The equations of motion can be obtained from the stationary action principle
by taking the variation of the functional
\begin{equation*}
S[x, \psi, \overline{\psi}]=\int\limits_{\tau_{1}}^{\tau_{2}} L(x, \psi, \overline{\psi})d\tau
\end{equation*}
with respect to the variables $z^{A}=(x^{\mu}, \psi, \overline{\psi})$
under the condition
\begin{equation*}
\left.\delta z^{A}(\tau)\right|_{\tau=\tau_{1, 2}}=0\,.
\end{equation*}

Consider the explicit variation of $L$ extracting a total time derivative.
Evidently, for the synchronic variations $\delta z^{A}$
(i. e. $\partial\delta z^{A}=\delta\partial z^{A}$)
it can be written as
\begin{equation}
\begin{split}
\delta L=&\eta_{\mu\nu}v^{\mu}\delta v^{\nu}=\partial \left[ p_{\mu} \left(\delta x^{\mu}+
\frac{i}{2}\left( \overline{\psi}\gamma^{\mu}\delta\psi -\delta\overline{\psi}\gamma^{\mu}\psi\right) \right)\right]\\
-&\partial p_{\mu}\delta x^{\mu}
+\frac{i}{2}\delta\overline{\psi}\left[p_{\mu}\gamma^{\mu}\partial\psi+\partial(p_{\mu}\gamma^{\mu}\psi)\right]\\
-&\frac{i}{2}\left[ \overline{\psi}\gamma^{\mu}p_{\mu}+\partial(\overline{\psi}\gamma^{\mu}p_{\mu}) \right]\delta\psi\\
\equiv&\partial\left[ p_{\mu}\left( \delta x^{\mu}+\frac{i}{2}\left( \overline{\psi}\gamma^{\mu}\delta\psi-\delta\overline{\psi}\gamma^{\mu}\psi \right)\right)\right]
-\delta x^{\mu}\partial p_{\mu}\\
+&\frac{i}{2}\left[ 2\hat{p}\partial\psi+\left( \partial\hat{p} \right)\psi\right]
-\frac{i}{2}\left[ 2\partial\overline{\psi}\hat{p}+\overline{\psi}\partial\hat{p} \right]\,,
\end{split}
\label{2-3}
\end{equation}
where we denote
\begin{equation}
p_{\mu}=\eta_{\mu\nu}v^{\nu}\,,\quad
\hat{p}=p_{\mu}\gamma^{\mu}\,.
\label{2-4}
\end{equation}
Taking into account the arbitrariness of the variations $\delta z^{A}$
one can obtain the equations of motion in the following form
\begin{equation}
\partial p_{\mu}=0\,,\quad
\hat{p}\partial\psi+\frac{1}{2}(\partial\hat{p})\psi=0\,,\quad
(\partial\overline{\psi})\hat{p}+\frac{1}{2}\overline{\psi}\partial\hat{p}=0\,.
\label{2-5}
\end{equation}

Making a substitution of the first equation into the other ones
we get
\begin{equation}
\partial p_{\mu}=0\,,\quad
\hat{p}\partial\psi=0\,,\quad
\partial\overline{\psi}\hat{p}=0\,.
\label{2-6}
\end{equation}
The same result can be obtained from the Euler-Lagrange equations
which have to be used with a certain accuracy because of the presence of
Grassman variables. A little complication arises due to two different ways
of the definition of derivation with respect to $\psi$ and
$\overline{\psi}$ (left and right derivatives). These definitions appear
when the factor $\delta z$ is put outside the brackets in the variation
of an arbitrary function $F(z)$ by two different ways, namely
\begin{equation}
\delta F(z)=F(z+\delta z)-F(z)=\delta z^{A}\frac{\vec{\partial} F}{\vec{\partial} z^{A}}=
\frac{\overleftarrow{\partial} F}{\overleftarrow{\partial} z^{A}}\delta z^{A}\,,
\label{2-7}
\end{equation}
here the symbols $\vec{\partial}$ and $\overleftarrow{\partial}$
denote the left and right derivation correspondingly,
and
\begin{gather*}
\frac{\vec{\partial}}{\vec{\partial} z}\left( f_{1}(z)f_{2}(z) \right)=
\frac{\vec{\partial} f_{1}(z)}{\vec{\partial} z}f_{2}(z)+
(-1)^{\mathfrak{a}(f_{1})}f_{1}
\frac{\vec{\partial}f_{2}}{\vec{\partial}z}\,,\\[2mm]
\frac{\overleftarrow{\partial}}{\overleftarrow{\partial} z}\left( f_{1}(z)f_{2}(z) \right)=
(-1)^{\mathfrak{a}(f_{2})}
\frac{\overleftarrow{\partial} f_{1}(z)}{\overleftarrow{\partial} z}f_{2}(z)+
f_{1}\frac{\overleftarrow{\partial}f_{2}}{\overleftarrow{\partial}z}
\end{gather*}
It is easy to prove that
\begin{equation*}
\frac{\vec{\partial}F}{\vec{\partial} z}=
-(-1)^{\mathfrak{a}(F)}
\frac{\overleftarrow{\partial} F}{\overleftarrow{\partial} z}\,.
\end{equation*}

The derivatives with respect to odd variables are anticommutative \cite{berezin}.
We assume that the derivative with respect to $\psi$ and $\overline{\psi}$
hereinafter mean the right and left-side derivation respectively.

Generalized momenta conjugated to $x^{\mu}$, $\psi$, $\overline{\psi}$
are
\begin{subequations}
\label{2-8}
\begin{align}
\mathcal{P}_{\mu}&=\frac{\partial L}{\partial \partial x^{\mu}}=\eta_{\mu\nu}v^{\mu}\equiv p_{\mu}\,,\\
\mathcal{P}_{\psi}&=\frac{\partial L}{\partial \partial \psi}=\frac{i}{2}\overline{\psi}\hat{p}\,,\\
\mathcal{P}_{\overline{\psi}}&=\frac{\partial L}{\partial \partial \overline{\psi}}=-\frac{i}{2}\hat{p}\psi\,.
\end{align}
\end{subequations}
A direct calculation shows that the Hamiltonian coincides with the
Lagrangian. This fact means that the theory under consideration is
purely kinematic.

The Lagrangian after some algebraic calculations can be rewritten in the
following form
\begin{equation}
L=\frac{1}{2}\partial z^{A} g_{AB}(z)\partial z^{B}\,,
\label{2-9}
\end{equation}
the structure of the ``metric'' $g_{AB}$ can be obtained from (\ref{2-2})
after the direct substitution of (\ref{2-1})
\begin{equation*}
\begin{split}
&\{g_{AB}\}=\\
&
\begin{pmatrix}
\eta_{\mu\nu}&\dfrac{i}{2}(\overline{\psi}\gamma_{\mu})_{a}&\dfrac{i}{2}(\gamma_{\mu}\psi)^{a}\\[2mm]
\dfrac{i}{2}(\overline{\psi}\gamma_{\mu})_{a}&\dfrac{1}{4}(\overline{\psi}\gamma^{\mu})_{a}\eta_{\mu\nu}(\overline{\psi}\gamma^{\nu})_{b}&
\dfrac{1}{4}(\gamma^{\mu}\psi)^{a}\eta_{\mu\nu}(\overline{\psi}\gamma^{\nu})_{b}\\[4mm]
-\dfrac{i}{4}(\gamma_{\mu}\psi)^{a}&\dfrac{1}{4}(\overline{\psi}\gamma^{\mu})_{a}\eta_{\mu\nu}(\gamma^{\nu}\psi)^{b}&
\dfrac{1}{4}(\gamma^{\mu}\psi)^{a}\eta_{\mu\nu}(\gamma^{\nu}\psi)^{b}
\end{pmatrix}
\end{split}
\end{equation*}
One can arrive to the same result after two-fold differentiation of the
Lagrangian with respect to the velocities taking into account the side of
derivation.

Note that there are the following equivalent form of the Lagrangian
that can be obtained with the use of the definitions
(\ref{1-12}), (\ref{1-13}),
\begin{equation*}
\begin{split}
L&=\frac{1}{2}v^{\mu}\eta_{\mu\nu}v^{\nu}=
\frac{1}{2}(E^{\mu}_{A}\partial z^{A})\eta_{\mu\nu}(E^{\nu}_{B}\partial z^{B})\\
&=\frac{1}{2}s(E^{\mu}_{A}, z^{A}) \partial z^{A}\left( E^{\mu}_{A}\eta_{\mu\nu}E^{\nu}_{B} \right)\partial z^{B}\,.
\end{split}
\end{equation*}


\section{Classical Symmetries of Lagrangian for Superparticle}
\label{s3}

The infinitesimal coordinate transformation
$z^{A}\to z^{\prime A}=z^{A}+\delta z^{A}$ in the superspace is a symmetry of
the dynamical theory when $\delta L$ is equal to zero or can be presented as
a total time derivative of some function. In the case of mechanical systems
it is possible to redefine the Lagrangian mainaining the character of a theory
in order to achieve $\delta L=0$. Since we assume that such a
property of the Lagrangian holds.

Taking into account the definition of the Lagrangian we can write in the
case of synchronic variations ($\partial\delta z=\delta\partial z$)
\begin{equation*}
\begin{split}
\delta L&=\frac{1}{2}\left[ \partial\delta z^{A}g_{AB}\partial z^{B}+
\partial z^{A}g_{AB}\partial\delta z^{B} \right.\\
&+\left.\partial z^{A}\left(\frac{\partial g_{AB}}{\partial z^{C}}\delta z^{C}\right)\partial z^{B} \right]\,.
\end{split}
\end{equation*}
Since $\delta z^{A}=\delta z^{A}(z)$, then $\delta\partial z^{A}=\partial\delta z^{A}=\delta z^{C}\dfrac{\partial\delta z^{A}}{\partial z^{C}}$,
so that
\begin{equation}
\begin{split}
\delta L=\partial z^{A}\left[ g_{CB}\partial_{A}\delta z^{C}\right.&+\left.g_{AC}\partial_{B}\delta z^{C}\right.\\
&+\left.\delta z^{C}\partial_{C}g_{AB} \right]\partial z^{B}=0\,.
\end{split}
\label{3-1}
\end{equation}
Hence the symmetries of the Lagrangian are isometries of the metric
$\{g_{AB}\}$ and the variations $\delta z^{A}$ obey Killing equations.

In order to write down the explicit form of the Killing equations
it is suitable to calculate directly the variation of $L$
under the transformation $z\to z'=z+\delta z(z)$. Using (\ref{2-1}) we find
\begin{equation}
\delta L=v^{\mu}\eta_{\mu\nu}\delta v^{\nu}\equiv p_{\mu}\delta v^{\mu}\,.
\label{3-2}
\end{equation}
It is not difficult to express $\delta v^{\mu}$
in terms of the variations
$\delta x^{\mu}(x, \psi, \overline{\psi})$, $\delta \psi(x, \psi, \overline{\psi})$
and $\delta \overline{\psi}(x, \psi, \overline{\psi})$
taking into account the properties of derivation operations
\begin{equation}
\begin{split}
\delta v^{\mu}&=
\partial x^{\nu}\left[ \frac{\partial \delta x^{\mu}}{\partial x^{\nu}}+
\frac{i}{2}\left( \overline{\psi}\gamma^{\mu}\frac{\partial\delta\psi}{\partial x^{\nu}}-\frac{\partial \delta\overline{\psi}}{\partial x^{\nu}}\gamma^{\mu}\psi \right)\right]\\
&+
\partial\overline{\psi}
\left[ \frac{\partial \delta x^{\mu}}{\partial \overline{\psi}}-
\frac{i}{2}\left( \gamma^{\mu}\delta\psi+\overline{\psi}\gamma^{\mu}\frac{\partial \delta\psi}{\partial \overline{\psi}}+
\frac{\partial \delta\overline{\psi}}{\partial \overline{\psi}}\gamma^{\mu}\psi \right)\right]\\
&+
\left[ \frac{\partial \delta x^{\mu}}{\partial \psi}+\frac{i}{2}\left(\delta\overline{\psi}\gamma^{\mu}+
\overline{\psi}\gamma^{\mu}\frac{\partial \delta\psi}{\partial \psi}+
\frac{\partial \delta\overline{\psi}}{\partial \psi}\gamma^{\mu}\psi  \right)\right]\partial\psi\,.
\label{3-3}
\end{split}
\end{equation}
Let us rewrite (\ref{3-2}) using (\ref{3-3}) as following
\begin{equation}
\delta L=v^{\mu}\left( A_{\mu\nu}\partial x^{\nu}+\partial\overline{\psi}B_{\mu}+\overline{B}_{\mu}\partial\psi \right)\,,
\label{3-4}
\end{equation}
where the coefficients multiplying on the velocities $\partial x^{\mu}$,
$\partial\psi$ and $\partial\overline{\psi}$ are denoted as
\begin{subequations}
\label{3-5}
\begin{align}
A_{\mu\nu}&=\frac{\partial \delta x_{\mu}}{\partial x^{\nu}}+
\frac{i}{2}\left( \overline{\psi}\gamma_{\mu}\frac{\partial\delta\psi}{\partial x^{\nu}}-\frac{\partial
\delta\overline{\psi}}{\partial x^{\nu}}\gamma_{\mu}\psi \right)\,,\\
B_{\mu}&=\frac{\partial \delta x_{\mu}}{\partial \overline{\psi}}-
\frac{i}{2}\left( \gamma_{\mu}\delta\psi+\overline{\psi}\gamma_{\mu}\frac{\partial \delta\psi}{\partial \overline{\psi}}+
\frac{\partial \delta\overline{\psi}}{\partial \overline{\psi}}\gamma_{\mu}\psi \right)\,,\\
\overline{B}_{\mu}&=\frac{\partial \delta x_{\mu}}{\partial \psi}+\frac{i}{2}\left(\delta\overline{\psi}\gamma_{\mu}+
\overline{\psi}\gamma_{\mu}\frac{\partial \delta\psi}{\partial \psi}+
\frac{\partial \delta\overline{\psi}}{\partial \psi}\gamma_{\mu}\psi  \right)
\end{align}
\end{subequations}
(here the lowering and raising index operations with the Greek indices
$\mu, \nu \dots$ are defined in the metric $\eta_{\mu\nu}$).

Using the definition (\ref{2-1}) of the supervelocities $v^{\mu}$
we can write the final expression for $\delta L$
\begin{equation*}
\begin{split}
\delta L&=
\left[ \partial x^{\mu}+\frac{i}{2}\left(
\overline{\psi}\gamma^{\mu}\partial\psi
-\partial\overline{\psi}\gamma^{\mu}\psi \right) \right]\\
&\times
\left[ A_{\mu\nu}\partial x^{\nu}+\partial\overline{\psi}B_{\mu}+\overline{B}_{\mu}\partial\psi \right]\,,
\end{split}
\end{equation*}
and, after comparing the factors corresponding to the different products of
$\partial z^{A}$, we obtain the following Killing equations
\begin{subequations}
\label{3-6}
\begin{gather}
A_{\mu\nu}+A_{\nu\mu}=0\\
\overline{B}_{\nu}+\frac{i}{2}\overline{\psi}\gamma^{\mu}A_{\mu\nu}=0\,,\quad
B_{\nu}-\frac{i}{2}\gamma^{\mu}\psi A_{\mu\nu}=0\,,\\[1mm]
(B_{\mu})_{b}(\overline{\psi}\gamma^{\mu})^{a}-(\gamma^{\mu}\psi)_{b}(\overline{B})^{a}=0\,,\\[1mm]
(\overline{\psi}\gamma^{\mu})_{a}(\overline{B}_{\mu})_{b}-(\overline{\psi}\gamma^{\mu})_{b}(\overline{B}_{\mu})_{a}=0\,,\\
(\gamma^{\mu}\psi)^{a}(B_{\mu})^{b}-(\gamma^{\mu}\psi)^{b}(B_{\mu})^{a}=0\,.
\end{gather}
\end{subequations}

The Killing equations (\ref{3-6}) are partial differential equations
of first order. In the case of the variations determined by the following
functional structure
$\delta\psi=\delta\psi(\psi)$, $\delta\overline{\psi}=\delta\overline{\psi}(\overline{\psi})$
these equations become more simple ones and the objects (\ref{3-5}) are reduced to
\begin{gather*}
A_{\mu\nu}=\frac{\partial \delta x_{\mu}}{\partial x^{\nu}}\,,\quad
B^{\mu}=\frac{\partial \delta x^{\mu}}{\partial \overline{\psi}}-
\frac{i}{2}\left( \gamma^{\mu}\delta\psi+\frac{\partial \delta\overline{\psi}}{\partial \overline{\psi}}\gamma^{\mu}\psi
\right)\,,\quad\\
\overline{B}^{\mu}=\frac{\partial \delta x^{\mu}}{\partial \psi}+
\frac{i}{2}\left( \delta\overline{\psi}\gamma^{\mu}+\overline{\psi}\gamma^{\mu}\frac{\partial \delta\psi}{\partial \psi} \right)\,.
\end{gather*}

The Killing equations for the spatial coordinates $x^{\mu}$ read
\begin{equation*}
\frac{\partial\delta x_{\mu}}{\partial x^{\nu}}+\frac{\partial\delta x_{\nu}}{\partial x^{\mu}}=0\,.
\end{equation*}
Its solutions independent on $\psi$ and $\overline{\psi}$ are well-known,
they describe the representation of the Poincare group $\iso{p}{q}$, $p+q=n$:
\begin{equation}
\delta x^{\mu}=\varepsilon^{\mu}+\varepsilon^{\mu}{}_{\nu}x^{\nu}=
\varepsilon^{\mu}+\frac{1}{2}\varepsilon^{\alpha\beta}(\delta^{\mu}_{\alpha}\eta_{\nu\beta}-\delta^{\mu}_{\beta}\eta_{\nu\alpha})x^{\nu}\,,
\label{3-7}
\end{equation}
here $\varepsilon^{\mu}$ are infinitesimal parameters of the translation
subgroup $\T{n}$ and the infinitesimal parameters
$\varepsilon^{\mu\nu}=-\varepsilon^{\nu\mu}$ are related to the Lorentz
transformation group $\so{p}{q}$.
It is easy to find the solution for the variations of the spinor coordinates
$\psi$ and $\overline{\psi}$,
\begin{equation}
\delta\psi=-\frac{i}{2}\hat{\varepsilon}\psi\,,\quad
\delta\overline{\psi}=\frac{i}{2}\overline{\psi}\hat{\varepsilon}\,,
\label{3-8}
\end{equation}
where
\begin{equation*}
\hat{\varepsilon}=\varepsilon_{\mu\nu}\sigma^{\mu\nu}\,,\quad
\sigma^{\mu\nu}=\frac{i}{4}[\gamma^{\mu}, \gamma^{\nu}]\,.
\end{equation*}
In the case of these variations the functions (\ref{3-5}) are reduced to
\begin{equation*}
A_{\mu\nu}=\varepsilon_{\mu\nu}\,,\quad
B^{\mu}=\frac{1}{4}[\gamma^{\mu}, \hat{\varepsilon}]\psi\,,\quad
\overline{B}^{\mu}=-\frac{1}{4}\overline{\psi}[\gamma^{\mu}, \hat{\varepsilon}]\,.
\end{equation*}
Taking into account the property of Dirac matrices
$[\sigma^{\alpha\beta}, \gamma^{\mu}]=i(\gamma^{\alpha}\eta^{\mu\beta}-\gamma^{\beta}\eta^{\mu\alpha})$,
one can observe that
\begin{equation*}
[\hat{\varepsilon}, \gamma^{\mu}]=-2i\varepsilon^{\mu}{}_{\alpha}\gamma^{\alpha}\,,
\end{equation*}
then
\begin{equation*}
\delta v^{\mu}=\varepsilon^{\mu}{}_{\alpha}v^{\alpha}\,.
\end{equation*}
Hence, using (\ref{3-2}), one can draw a conclusion that $\delta L=0$
for the transformation with the parameters
$\varepsilon^{\mu}$ and $\varepsilon^{\mu\nu}$.

On the other hand, with the same conditions, there exists the solution of
the Killing equations with the functional structure
$\delta x^{\mu}=\delta x^{\mu}(\psi, \overline{\psi})$, which has the
following form
\begin{equation*}
\delta x^{\mu}=\frac{i}{2}\left( \overline{\psi}\gamma^{\mu}\xi-\overline{\xi}\gamma^{\mu}\psi \right)\,,\quad
\delta \psi=\xi\,,\quad \delta\overline{\psi}=\overline{\xi}\,,
\end{equation*}
where $\xi$, $\overline{\xi}$ are infinitesimal spinor parameters.
For these transformations we have
$B^{\mu}=0$, $\overline{B}^{\mu}=0$, $A_{\mu\nu}=0$,
therefore $\delta v^{\mu}=0$ and $\delta L\equiv 0$.
Such a transformation describes the supertranslations with the parameters
$\xi$ and $\overline{\xi}$.

It should be noted here that in principle the Killing equations
can possess the more wide class of solutions than described above,
but in our consideration it is not essential.


\section{Lagrangian in Quantum Theory}
\label{s4}

In quantum theory the coordinates $\{x^{\mu}, \psi, \overline{\psi}\}$
become operators that are, generally  speaking, non-commutative objects.
Before writing down the Lagrangian and other standard constructions explicitly
the manner of operator ordering in products have to be decided.
Let us introduce the generalization of a Jordan product for the case of the
presence of odd variables
\begin{equation}
a\circ b:=\frac{1}{2}(ab+s(a, b)ba)\,,
\label{4-1}
\end{equation}
where $s(a, b)$ is the signature factor for the operators $a$ and $b$
defined in (\ref{1-1}).

The velocity operator can be defined as
\begin{equation}
v^{\mu}=\partial x^{\mu}+\frac{i}{2}\left( \overline{\psi}\gamma^{\mu}\circ\partial\psi-
\partial\overline{\psi}\gamma^{\mu}\circ\psi\right)\,.
\label{4-2}
\end{equation}

As to the operator properties of dynamical variables we assume that
the momentum operator $p_{\mu}$ commutes with $\psi$ and $\overline{\psi}$
(this assumption will be confirmed below). Then the quantum Lagrangian
receives the form
\begin{equation}
L=\frac{1}{2}v^{\mu}\eta_{\mu\nu}v^{\nu}\,.
\label{4-3}
\end{equation}
This expression is invariant under Poincare group transformations.

The variation of (\ref{4-3}) under arbitrary synchronic coordinate
transformation $z^{A}\to z'{}^{a}=z^{A}+\delta z^{A}$ reads
\begin{equation}
\begin{split}
\delta L&=p_{\mu}\circ\delta v^{\mu}=\partial \mathcal{Q}\\
&+\frac{i}{2}\left[ \delta \overline{\psi}\circ\left(
\hat{p}\partial\psi+\partial (\hat{p}\psi) \right)- \left(
\partial\overline{\psi}\hat{p}+\partial(\overline{\psi}\hat{p})
\right) \right]\\
&-\partial p_{\mu}\circ\delta x^{\mu}\,,
\end{split}
\label{4-4}
\end{equation}
where
\begin{equation}
\mathcal{Q}=p_{\mu}\circ\delta x^{\mu}+\frac{i}{2}\left( \overline{\psi}\hat{p}\circ\delta\psi-\delta\overline{\psi}\hat{p}\circ\psi \right)\,,\quad
\hat{p}=p_{\mu}\gamma^{\mu}\,.
\label{4-5}
\end{equation}
Here we have used the assumptions $[p_{\mu}, \psi]=0$, $[p_{\mu},
\overline{\psi}]=0$, so that (\ref{4-4}) holds for any operator
properties of $\delta z$.

When the variation ``$\delta$'' is related to permissible variations,
i.~e.~$\delta L=0$ (because $L$ is purely kinematic object),
the dynamical equations obtain the form
\begin{equation}
\partial p_{\mu}=0\,,\quad
\hat{p}\partial\psi=0\,,\quad
\partial\overline{\psi}\hat{p}=0\,.
\label{4-6}
\end{equation}
In addition, in the case of permissible variations there is the following
conservation law
\begin{equation*}
\partial \mathcal{Q}=0\,.
\end{equation*}
This means that $\mathcal{Q}$ is the generator of permissible variations.

In quantum theory the permissible variations determined by $\delta L=0$
without any referring to equations of motion are described by
the properties that are different from the language of classical
Killing equations (we cannot give the
rigorous definition of the derivatives with respect to non-commutative
operators $x^{\mu}$, $\psi$, $\overline{\psi}$). Nevertheless,
a direct calculation shows that supertranslations and $\so{p}{q}$-transformations
are permissible variations. Let us find the explicit expressions for
the generators of these transformations.

In the case of the supertranslations one can write
\begin{equation}
\delta x^{\mu}=\varepsilon^{\mu}+\frac{i}{2}\left( \overline{\psi}\gamma^{\mu}\xi-\overline{\xi}\gamma^{\mu}\psi\right)\,,\quad
\delta\psi=\xi\,,\quad \delta\overline{\psi}=\overline{\xi}
\label{4-7}
\end{equation}
and the generator reads
\begin{equation}
G=\varepsilon^{\mu}p_{\mu}+i(\overline{\psi}\hat{p}\xi-\overline{\xi}\hat{p}\psi)
\label{4-8}
\end{equation}
(as it follows from (\ref{4-5})).

Similarly, taking the variation related to $\so{p}{q}$-trans\-for\-mations
one obtains
\begin{equation}
\delta x^{\mu}=\varepsilon^{\mu}{}_{\nu}x^{\nu}\,,\quad
\delta\psi=-\frac{i}{2}\hat{\varepsilon}\psi\,,\quad
\delta\overline{\psi}=\frac{i}{2}\overline{\psi}\hat{\varepsilon}
\label{4-9}
\end{equation}
with
\begin{equation}
G=-\frac{1}{2}\varepsilon^{\mu\nu}J_{\mu\nu}\,,\quad
J_{\mu\nu}=L_{\mu\nu}+S_{\mu\nu}\,,
\label{4-10}
\end{equation}
where
\begin{equation}
L_{\mu\nu}=x_{\mu}p_{\nu}-x_{\nu}p_{\mu}\,,\quad
S_{\mu\nu}=-\frac{1}{2}\overline{\psi}\{\hat{p}, \sigma_{\mu\nu}\}\psi\,.
\label{4-10a}
\end{equation}
Here we assume that $[x_{\mu}, p_{\nu}]=[x_{\nu}, p_{\mu}]$, therefore
the symmetrization in the expression of $L_{\mu\nu}$ is unnecessary.
The curly brackets denote the anticommutator of {\it matrices}.
In our context the operators $L_{\mu\nu}$ and $S_{\mu\nu}$
have the sense of orbital and spin momentae of a superparticle.

Basing on (\ref{4-8}) and (\ref{4-10}) it is possible to
develop the algebra of commutation relations, this analyzes
will be carried out in the next section.


\section{Commutation Relations and Charges}
\label{s5}

If $G$ denotes the generator of permissible variations then the variation
of an arbitrary operator $\mathcal{F}$ obeys the equation
\begin{equation}
\delta\mathcal{F}=\frac{1}{i\hbar}[\mathcal{F}, G]\,,
\label{5-1}
\end{equation}
When $\mathcal{F}$ and $G$ are given and the algebra of commutation relations
is known, the variation $\delta\mathcal{F}$ can be directly calculated
from~(\ref{5-1}). In our context we have the explicit forms for $\mathcal{F}$
and $G$ (determined from the analyzes of the symmetries of $L$)
and~(\ref{5-1}) allow to obtain some information about the operator
algebra using the properties (given a priory) of the system  under
consideration.

Making the substitutions of different operators $\mathcal{F}$ with the
same generator $G$ one arrive to the algebraic system for unknown
commutators which is totally or partially solvalable depending on
character of a model.

In order to construct the algebra of commutation relations let us
consider supertranslations as the permissible variations. The corresponding
generator can be presented as
\begin{equation}
G=p_{\mu}\varepsilon^{\mu}+\overline{Q}\xi+\overline{\xi}Q\,,
\label{5-2}
\end{equation}
where
\begin{equation*}
\overline{Q}=i\overline{\psi}\hat{p}\,,\quad
Q=-i\hat{p}\psi\,.
\end{equation*}
Using the definition of the variations (\ref{4-7}) we can obtain after
the simple calculation the following commutation relations for
spinor coordinates
\begin{equation}
\begin{gathered}
\mbox{} [\psi, p_{\mu}]=0\,,\quad
[\psi, \overline{\psi}\hat{p}]=\hbar\1\,,\quad
[\psi, \hat{p}\psi]=0\,,\\
[\overline{\psi}, p_{\mu}]=0\,,\quad [\overline{\psi}, \hat{p}\psi]=\hbar\1\,,\quad
[\overline{\psi}, \overline{\psi}\hat{p}]=0\,,
\label{5-3}
\end{gathered}
\end{equation}

where $\1$ denotes the unit matrix with spinor indices.
Analogously, for the coordinates of $\M^{n}$ we can write
\begin{equation}
\begin{gathered}
\mbox{} [x^{\mu}, p_{\nu}]=i\hbar\delta^{\mu}_{\nu}\,,\\
[x^{\mu}, \overline{\psi}\hat{p}]=\frac{i\hbar}{2}\overline{\psi}\gamma^{\mu}\,,\quad
[x^{\mu}, \hat{p}\psi]=\frac{i\hbar}{2}\gamma^{\mu}\psi\,.
\end{gathered}
\label{5-4}
\end{equation}
Note that (\ref{5-3}) is consistent with the initial assumption about
the commutativity of the momentum $p_{\mu}$ with  the operators
$\psi$ and $\overline{\psi}$.

Putting $p_{\mu}$ instead of $\mathcal{F}$ in the formula~(\ref{5-1})
we obtain
\begin{equation}
[p_{\mu}, p_{\nu}]=0\,,\quad
[p_{\mu}, \overline{\psi}\hat{p}]=0\,,\quad
[p_{\mu}, \hat{p}\psi]=0\,.
\label{5-5}
\end{equation}
Taking into account the commutativity of momenta $p_{\mu}$,
define the inverse to $\hat{p}$ operator $\hat{\alpha}$ as
\begin{equation*}
\hat{p}\hat{\alpha}=\hat{\alpha}\hat{p}=\1\,,
\end{equation*}
where the property of commutativity is related to the matrix elements\footnote{%
i. e.
$(\hat{\alpha})^{a}_{b}(\hat{p})^{b}_{c}=(\hat{p})^{b}_{c}(\hat{\alpha})^{a}_{b}=\delta^{a}_{c}$}.
It is possible to give the explicit expression for $\hat{\alpha}$.
Evidently, $\hat{p}\hat{p}=p_{\mu}p_{\nu}\gamma^{\mu}\gamma^{\nu}=\eta^{\mu\nu}p_{\mu}p_{\nu}$,
So that
\begin{equation*}
\hat{\alpha}=\frac{\hat{p}}{\eta^{\mu\nu}p_{\mu}p_{\nu}}:=\alpha_{\mu}\gamma^{\mu},\quad
\alpha_{\mu}=\frac{p_{\mu}}{\eta^{\mu\nu}p_{\mu}p_{\nu}}\,.
\end{equation*}

Then, using the formulae (\ref{5-3})--(\ref{5-5}) we can write
\begin{subequations}
\label{5-6}
\begin{align}
&[\psi, \psi]=0\,,\quad [\overline{\psi}, \overline{\psi}]=0\,,\quad
[\psi, \overline{\psi}]=\hbar\1\,,\\
&[\psi, p_{\mu}]=0\,,\quad [\overline{\psi}, p_{\mu}]=0\,,\\
&[x^{\mu}, p_{\nu}]=i\hbar\delta^{\mu}_{\nu}\,,\quad [p_{\mu}, p_{\nu}]=0\,,\\
&[x^{\mu}, \psi]=-\frac{i\hbar}{2}\hat{\alpha}\gamma^{\mu}\psi\,,\quad
[x^{\mu}, \overline{\psi}]=-\frac{i\hbar}{2}\overline{\psi}\gamma^{\mu}\hat{\alpha}\,.
\end{align}
\end{subequations}
It is useful to write down the commutators between $x^{\mu}$ and
the operators $\hat{p}$ and $\hat{\alpha}$, which can be calculated
immediately from the definitions of these objects
\begin{equation}
[x^{\mu}, \hat{p}]=i\hbar\gamma^{\mu}\,,\quad
[x^{\mu}, \hat{\alpha}]=-i\hbar\hat{\alpha}\gamma^{\mu}\hat{\alpha}\,.
\label{5-7}
\end{equation}

The commutator $[x^{\mu}, x^{\nu}]$ can be obtained by indirect way
with the use of Jacobi identity for the triple of operators $x^{\mu}$, $x^{\nu}$
and $\psi$. It is easy to prove that
\begin{equation}
[x^{\mu}, x^{\nu}]=\frac{\hbar}{4}\overline{\psi}(\gamma^{\mu}\hat{\alpha}\gamma^{\nu}-
\gamma^{\nu}\hat{\alpha}\gamma^{\mu})\circ\psi\,.
\label{5-8}
\end{equation}
Using the properties of $\gamma$-matrices one can rewrite~(\ref{5-8}) as
\begin{equation}
[x^{\mu}, x^{\nu}]=\frac{i\hbar}{2}\overline{\psi}\{\hat{\alpha}, \sigma^{\mu\nu}\}\circ\psi\,,
\label{5-9}
\end{equation}
where the curly brackets are related to the anticommutator of matrices.

The commutation relations (\ref{5-9}) show that in quantum theory of
a superparticle the coordinates $x^{\mu}$, $\psi$ and $\overline{\psi}$
are not commutative with each other. Therefore the features
of quantum theory on a superspace essentially differ from the
corresponding theory on a usual Riemannian manifold. The explicit
manifestation of them consists in non-commutativity of geometrical
coordinates $\{x^{\mu}\}$ related to the underlying manifold $\M^{n}$.

Now, using obtained above commutative relations, let us consider the
algebra of charges describing the $\so{p}{q}$ symmetry.
\begin{equation*}
\delta x^{\mu}=\varepsilon^{\mu}{}_{\nu}x^{\nu}\,,\quad
\delta\psi=-\frac{i}{2}\hat{\varepsilon}\psi\,,\quad
\delta\overline{\psi}=\frac{i}{2}\overline{\psi}\hat{\varepsilon}\,.
\end{equation*}
The generator of these transformations has the form
\begin{equation}
G=-\frac{1}{2}\varepsilon^{\mu\nu}J_{\mu\nu}\,,
\label{5-10}
\end{equation}
where
\begin{gather*}
J_{\mu\nu}=L_{\mu\nu}+S_{\mu\nu}\,,\quad
L_{\mu\nu}=x_{\mu}p_{\nu}-x_{\nu}p_{\mu}\,,\\
S_{\mu\nu}=-\frac{1}{2}\overline{\psi}\{\hat{p}, \sigma_{\mu\nu}\}\circ\psi\,.
\end{gather*}

Then, making a substitution of the variations of charges describing
supertranslations and $\so{p}{q}$-rotations to the main variational equation
(\ref{5-1}) with a total generator including these transformations,
one get the following algebra of charges
\begin{subequations}
\label{5-11}
\begin{gather}
 [p_{\mu}, p_{\nu}]=0\,,\quad [p_{\mu}, Q]=0\,,\quad [p_{\mu}, \overline{Q}]=0 \,,\\
 [p_{\mu}, J_{\alpha\beta}]=i\hbar(p_{\alpha}\eta_{\mu\beta}-p_{\beta}\eta_{\mu\alpha})\,,\\
 [Q, \overline{Q}]=\hbar\hat{p}\,,\quad
 [\overline{Q}, J_{\mu\nu}]=i\hbar\overline{Q}\sigma_{\mu\nu}\,,\\
 [Q, J_{\mu\nu}]=-i\hbar\sigma_{\mu\nu}Q\,,\\
\begin{aligned}
\mbox{}[J_{\mu\nu}, J_{\alpha\beta}]&= \\
i\hbar(\eta_{\mu\beta}&J_{\nu\alpha}+\eta_{\mu\alpha}J_{\mu\beta}-\eta_{\nu\beta}J_{\mu\alpha}-\eta_{\mu\alpha}J_{\nu\beta})\,.
\end{aligned}
\end{gather}
\end{subequations}
The algebra (\ref{5-11}) coincides with the usual algebra of
supersymmetrical extension of inhomogeneous Lorentz group
that has been introduced in \cite{volkov}--\cite{wess} in somewhat
different notations. It should be noted here, that because of
$[x^{\mu}, \psi]\ne 0$, $[x^{\mu}, \overline{\psi}]\ne 0$,
$[x^{\mu}, x^{\nu}]\ne 0$ the operators $L_{\mu\nu}$ and $S_{\mu\nu}$
considered separately do not form the Lie algebra of the form (\ref{5-11})
and due to this reason have not independent physical meaning.
As an integral of motion appears their combination $J_{\mu\nu}$.

In order to construct the coordinate representation and the space of states
it is necessarily to determine the basis of the Hilbert space of states $\H$.
As it well known, this basis is generated by a spectral problem of some
total set of commuting observables. In the superspace the coordinates
$\{z^{A}\}$ do not commute with each other and therefore
do not form a total set (as it holds in the usual quantum mechanics on the
Riemannian manifold). The total set can be constructed as a set of
combinations of the existing variables.

Define the operators
\begin{equation}
\eta^{\mu}_{\pm}=x^{\mu}\pm\frac{i}{2}\overline{\psi}\gamma^{\mu}\circ\psi\,.
\label{5-12}
\end{equation}
A direct calculation leads to the following commutation relations
\begin{subequations}
\label{5-13}
\begin{align}
[\eta^{\mu}_{+} ,\psi]&=-i\hbar\hat{\alpha}\gamma^{\mu}\psi\,,\quad
[\eta^{\mu}_{+}, \overline{\psi}]=0\,,\\
[\eta^{\mu}_{-}, \overline{\psi}]&=-i\hbar\overline{\psi}\gamma^{\mu}\hat{\alpha}\,, \quad
[\eta^{\mu}_{-}, \psi]=0\,,\\
[\eta^{\mu}_{\pm}, \eta^{\nu}_{\pm}]&=0\,,
\quad [\eta^{\mu}_{\pm}, \eta^{\nu}_{\mp}]=-[\eta^{\mu}_{\mp}, \eta^{\nu}_{\pm}]  \\
[\eta^{\mu}_{\pm}, \eta^{\nu}_{\mp}]&=\pm\frac{\hbar}{2}\overline{\psi}(\gamma^{\mu}\hat{\alpha}\gamma^{\nu}-\gamma^{\nu}\hat{\alpha}\gamma^{\mu})\circ\psi\,,
\end{align}
\end{subequations}
and
\begin{gather}
[\eta^{\mu}_{\pm}, p_{\nu}]=i\hbar\delta^{\mu}_{\nu}\,,\\
[\eta^{\mu}_{+},
x^{\nu}]=-\frac{\hbar}{2}\gamma^{\nu}\hat{\alpha}\gamma^{\mu}\circ\psi\,,\quad
[\eta^{\mu}_{-}, x^{\nu}]=\frac{\hbar}{2}\gamma^{\mu}\hat{\alpha}\gamma^{\nu}\circ\psi\,.
\label{5-14}
\end{gather}

These results can be obtained immediately from the main variational
equation with the generator of supertranslations, where one can put
\begin{equation*}
\delta\eta_{+}^{\mu}=\varepsilon^{\mu}+\overline{\psi}\gamma^{\mu}\xi\,,\quad
\delta\eta_{-}^{\mu}=\varepsilon^{\mu}-\overline{\xi}\gamma^{\mu}\psi\,.
\end{equation*}
The commutators between $\eta^{\mu}_{\pm}$ and $Q$, $\overline{Q}$ are
\begin{alignat*}{3}
[\eta^{\mu}_{+}, Q]&=0\,,\quad &
[\eta^{\mu}_{+}, \overline{Q}]&=i\hbar\overline{\psi}\gamma^{\mu}\,,\\
[\eta^{\mu}_{-}, Q]&=i\hbar\gamma^{\mu}\psi\,,\quad &
[\eta^{\mu}_{-}, \overline{Q}]&=0\,.
\end{alignat*}
One can observe that basing on the system of operators
$(x^{\mu}, \psi, \overline{\psi})$ it is possible to construct two
separated sets of
commuting observables $(\eta^{\mu}_{+}, \overline{\psi})$ and
$(\eta^{\mu}_{-}, \psi)$, which are not commutative with each other.

Note, that the velocity operator can be rewritten as
\begin{equation*}
v^{\mu}=\partial\eta^{\mu}_{+}-i\partial\overline{\psi}\gamma^{\mu}\circ\psi=
\partial\eta^{\mu}_{-}+i\overline{\psi}\gamma^{\mu}\circ\partial\psi\,.
\end{equation*}

Evidently, the derivatives of $\psi$ with respect to $\tau$
are not appear in the explicit expression of the Lagrangian presented
in terms of $(\eta^{\mu}_{+}, \overline{\psi})$, these derivatives are
included in the definition of momenta conjugated to $\overline{\psi}$
\begin{equation*}
\begin{split}
L=&\frac{1}{2}p_{\mu}\circ v^{\mu}=\frac{1}{2}\left[ \partial\eta^{\mu}_{+}\circ p_{\mu}+\partial\overline{\psi}\circ
(-i\hat{p}\psi)\right]\\
&\equiv\frac{1}{2}\left( \partial\eta^{\mu}_{+}\circ p_{\mu}+\partial\overline{\psi}\circ Q \right)\,.
\end{split}
\end{equation*}
It is easy to observe, that in such a definition the variables
$\psi$ have not corresponding conjugated momenta
(expressed as derivatives of $L$ with respect to $\partial\psi$
in the classical meaning). Therefore, the theory possesses constraints.


\section{Coordinate Representation}
\label{s6}

Let us choose the set of operators $\zeta=(\eta^{\mu}_{+}, \overline{\psi})$
as a total set of commutative observables, for the sake of abbreviation
we will write $\eta^{\mu}$ instead of $\eta^{\mu}_{+}$.
Define the basis states by the following way
\begin{equation*}
\zeta=|  \zeta' \rangle=\zeta'|  \zeta' \rangle\,,
\end{equation*}
or, more detally,
\begin{equation}
\eta^{\mu}|  \zeta' \rangle=\eta'{}^{\mu}|  \zeta' \rangle\,,\quad
\overline{\psi}|  \zeta' \rangle=\overline{\psi}'|  \zeta' \rangle\,.
\label{6-1}
\end{equation}
We choose the normalization condition in a standard form
\begin{equation}
\langle  \zeta' | \zeta''\rangle=\delta(\zeta'-\zeta'')
\label{6-2}
\end{equation}
and for an arbitrary function $f(\zeta')$ we can write
\begin{equation*}
\int f(\zeta'')\delta(\zeta'-\zeta'')d\zeta''=f(\zeta')\,,
\end{equation*}
where $\delta(\zeta'-\zeta'')=\delta(\eta'-\eta'')\delta(\overline{\psi}'-\overline{\psi}'')$.
In particular the $\delta$-function for spinor (Grassmanian) variables
reads
\begin{equation*}
\delta(\overline{\psi}'-\overline{\psi}'')=\overline{\psi}'-\overline{\psi}''
\end{equation*}
and $\delta(0)=1$ (differential and integral calculus for Grassmanian
numbers have been discussed in \cite{berezin}).

Now let us consider the matrix elements of the operators playing the important
role in the construction of the coordinate representation. Evidently,
in the case of operators which form the total set we can write
\begin{equation}
\langle  \zeta' |\overline{\psi}|  \zeta'' \rangle=\overline{\psi}''\langle  \zeta' |\zeta''\rangle\,,\quad
\langle  \zeta' |\eta^{\mu}|  \zeta'' \rangle=\eta''{}^{\mu}\langle  \zeta' |\zeta''\rangle\,.
\label{6-3}
\end{equation}
The matrix elements for momentum operators $p_{\mu}$ can be obtained
with the use of
\begin{equation*}
[\eta^{\mu}, p_{\nu}]=i\hbar\delta^{\mu}_{\nu}\,,\quad
[p_{\mu}, p_{\nu}]=0\,,
\end{equation*}
so, it is easy to find that
\begin{equation}
\langle  \zeta' |p_{\mu}|  \zeta'' \rangle=
-i\hbar\frac{\partial }{\partial \eta''{}^{\mu}}\langle  \zeta' |\zeta''\rangle+
\frac{\partial F(\zeta'')}{\partial \eta''{}^{\mu}}\langle  \zeta' |\zeta''\rangle\,,
\label{6-4}
\end{equation}
where $F(\zeta)$ is some continuous function. Analogously, from
\begin{equation*}
[\overline{\psi}, Q]=i\hbar\1\,,\quad
[\overline{\psi}, \overline{\psi}]=0\,,\quad
[Q, Q]=0
\end{equation*}
we get
\begin{equation}
\langle  \zeta' |Q|  \zeta'' \rangle=
-i\hbar\frac{\vec{\partial}}{\vec{\partial} \overline{\psi}''}\langle  \zeta' |\zeta''\rangle+
\frac{\vec{\partial} \Omega(\zeta'')}{\vec{\partial} \overline{\psi}''}\,,
\label{6-5}
\end{equation}
where we have taken into account the fact that
$\langle  \zeta' |\zeta''\rangle$ is an add number.
As in (\ref{6-4}), $\Omega(\zeta'')$ denotes some continuous function.
Hereinafter the derivation with respect to $\overline{\psi}$ is regarded
as a left one.

Since $[p_{\mu}, Q]=0$, then
\begin{equation}
\frac{\partial }{\partial \eta''{}^{\mu}}\frac{\partial }{\partial \overline{\psi}''}\Omega(\zeta'')=
\frac{\partial }{\partial \overline{\psi}''}\frac{\partial }{\partial \eta''{}^{\mu}}F(\zeta'')\,,
\label{6-6}
\end{equation}
and to a certain extent $\Omega$ can be identified with $F$
(they can differ on a linear function of $\zeta$ ).
Moreover, $\Omega$ and $F$ can be removed by a unitary transformation. So,
we can put $\Omega\equiv 0$, $F\equiv 0$.

Define the wave function for the state $|\Phi \rangle$ as a scalar
product
\begin{equation}
\Phi(\zeta)=\langle  \zeta | \Phi\rangle\,.
\label{6-7}
\end{equation}
Then the coordinate representation of the operators considered above
has the form
\begin{subequations}
\label{6-8}
\begin{align}
(\overline{\psi}\Phi)(\zeta')&=\overline{\psi}'\Phi(\zeta')\,,\quad
(\eta^{\mu}\Phi)(\zeta')=\eta'{}^{\mu}\Phi(\zeta')\,,\\
(p_{\mu}\Phi)(\zeta')&=-i\hbar\frac{\partial }{\partial \eta'{}^{\mu}}\Phi(\zeta')\,,\quad
(Q\Phi)(\zeta')=-i\hbar\frac{\partial }{\partial \overline{\psi}'}\Phi(\zeta')\,.
\end{align}
\end{subequations}

Using obtained results one can easily get the coordinate representation for
the generator of supertranslations on wave functions
\begin{equation*}
G=\varepsilon^{\mu}p_{\mu}+\overline{\xi}Q+\overline{Q}\xi
\end{equation*}
Since the variations of $\overline{\psi}$ and $\eta^{\mu}$ are
\begin{equation*}
\delta\overline{\psi}=\overline{\xi}\,,\quad
\delta\eta^{\mu}=\varepsilon^{\mu}+i\overline{\psi}\gamma^{\mu}\xi\,,
\end{equation*}
then the change of the wave function under such a variations of its
arguments reads
\begin{equation*}
\begin{split}
\delta\Phi(\zeta)&\equiv\delta\zeta\frac{\partial \Phi}{\partial \zeta}=
\delta\eta^{\mu}\frac{\partial \Phi}{\partial \eta^{\mu}}+
\delta\overline{\psi}\frac{\partial \Phi}{\partial\overline{\psi}}\\
&=(\varepsilon^{\mu}+i\overline{\psi}\gamma^{\mu}\xi)\frac{\partial \Phi}{\partial \eta^{\mu}}+
\overline{\xi}\frac{\partial \Phi}{\partial \overline{\psi}}\,.
\end{split}
\end{equation*}
On the other hand, according to the definition of the generator
we conclude that $\delta\Phi=\dfrac{i}{\hbar}G\Phi$, then taking into account
the coordinate representation of the operators $p_{\mu}$, $Q$ and $\overline{\psi}$
we arrive to the following expression
\begin{multline}
(G\Phi)(\zeta)=\\
-i\hbar\varepsilon^{\mu}\frac{\partial \Phi(\zeta)}{\partial \eta^{\mu}}
+i\overline{\psi}\left( -\gamma^{\mu}\frac{\partial }{\partial \eta^{\mu}} \right)\xi\Phi(\zeta)
-i\hbar\overline{\xi}\frac{\partial }{\partial \overline{\psi}}\Phi(\zeta)\\
=-i\hbar\left[ (\varepsilon^{\mu}+\overline{\psi}\gamma^{\mu}\xi)\frac{\partial }{\partial \eta^{\mu}}
+\overline{\xi}\frac{\partial }{\partial \overline{\psi}} \right]\Phi(\zeta)\,,
\end{multline}
that is in agreement with the previous conclusion.

Let us find the coordinate representation of other operators appearing in
commutation relations. In particular, consider the operator $\psi$.
Since
\begin{equation*}
[\overline{\psi}, p_{\mu}]=0\,,\quad [\psi, p_{\mu}]=0\,,
\end{equation*}
then
\begin{equation}
\frac{\partial }{\partial \eta''{}^{\mu}}\langle  \zeta' |\psi|  \zeta'' \rangle=
-\frac{\partial }{\partial \eta'{}^{\mu}}\langle  \zeta' |\psi|  \zeta'' \rangle
\label{6-9}
\end{equation}
(the same equality trivially holds also for $\overline{\psi}$ due to
$\langle  \zeta' |\overline{\psi}|\zeta''\rangle=\overline{\psi}''\langle\zeta' |\zeta''\rangle $).
In a similar way, from $[\psi, Q]=0$ we get
\begin{equation}
\frac{\partial }{\partial \overline{\psi}'}
\langle  \zeta' |\psi|  \zeta'' \rangle=
\frac{\partial }{\partial \overline{\psi}''}
\langle  \zeta' |\psi|  \zeta'' \rangle\,.
\label{6-10}
\end{equation}
Further, using the definition $Q=-i\hat{p}\psi$ we can write
\begin{equation*}
\begin{split}
-i\hbar\frac{\partial }{\partial \overline{\psi}''}\langle  \zeta' |\zeta''\rangle&\equiv
\langle  \zeta' |Q|  \zeta'' \rangle=
-i\int d\zeta'''\langle  \zeta' |\hat{p}|  \zeta''' \rangle\langle  \zeta''' |\psi|  \zeta'' \rangle\\
&=-\hbar\int d\zeta'''\left( \gamma^{\mu}\frac{\partial }{\partial \eta'''{}^{\mu}}\langle  \zeta' |\zeta'''\rangle \right)
\langle  \zeta''' |\psi|  \zeta'' \rangle\\
&=\hbar\gamma^{\mu}\frac{\partial }{\partial \eta'{}^{\mu}}\langle  \zeta' |\psi|  \zeta'' \rangle\,,
\end{split}
\end{equation*}
and, taking into account (\ref{6-9}) we conclude that the matrix element of
the operator $\psi$ is the solution of the differential equation
\begin{equation}
i\frac{\partial }{\partial \overline{\psi}''}\langle  \zeta' |\zeta''\rangle=
\gamma^{\mu}\frac{\partial }{\partial \eta''{}^{\mu}}\langle\zeta'|\psi|\zeta'' \rangle\,.
\label{6-11}
\end{equation}
This equality means that the operator $\psi$ is represented on wave
functions  by an integral operator. Multiplying (\ref{6-11}) on the
wave function and taking the integral of $\zeta$ (or $\zeta''$) we
find that
\begin{equation}
\frac{\partial }{\partial \overline{\psi}'}\Phi(\zeta')+
i\gamma^{\mu}\psi\frac{\partial }{\partial \eta'{}^{\mu}}\Phi(\zeta')=0
\label{6-12}
\end{equation}
(here the action of the operator $\psi$ precedes the differentiation).

The equation (\ref{6-12}) can be chosen as a definition of the realization of
the operator $\psi$, it can be rewritten as
\begin{equation}
(\psi\Phi)(\zeta)=-i[\hat{\partial}]^{-1}
\left( \frac{\partial }{\partial \overline{\psi}}\Phi(\zeta) \right)\,,
\label{6-13}
\end{equation}
where $[\hat{\partial}]^{-1}$ denotes the operator inversed to
$\hat{\partial}=\gamma^{\mu}\dfrac{\partial }{\partial \eta^{\mu}}$.

The action of the operator $\overline{Q}=i\overline{\psi}\hat{p}$
on wave functions is described by
\begin{equation}
(\overline{Q}\Phi)(\zeta')=-\left( \hbar\overline{\psi}'\gamma^{\mu}\frac{\partial }{\partial \eta'{}^{\mu}} \right)\Phi(\zeta')\,.
\label{6-14}
\end{equation}

To analyze the properties of the matrix elements of $\hat{\alpha}$
we take the matrix element of the equality $\hat{p}\hat{\alpha}=\hat{\alpha}\hat{p}=\1$,
using the representation of $p_{\mu}$ we get
\begin{equation*}
\langle  \zeta' |\hat{p}\circ \hat{\alpha}|  \zeta'' \rangle=
-\frac{i\hbar}{2}(\hat{\partial}''-\hat{\partial}')
\langle  \zeta' |\hat{\alpha}|  \zeta'' \rangle\equiv
\langle  \zeta' |\zeta''\rangle\,,
\end{equation*}
where the matrix indices of $\hat{\alpha}$ are contracted with the
matrix indices of $\gamma^{\mu}$.
The last equality shows that $\hat{\alpha}$ is realized by the
integral operator inversed to $\hat{\partial}$
\begin{equation*}
\hat{p}\Phi=-i\hbar\frac{\partial }{\partial \eta^{\mu}}\Phi\,,\quad
\hat{\alpha}\left( i\hbar\gamma^{\mu}\frac{\partial }{\partial \eta^{\mu}}\Phi \right)=-\Phi\,.
\end{equation*}
Since $[\overline{\psi}, \psi]=\hbar\hat{\alpha}$, it is easy to obtain the
equation connecting the matrix elements of  $\psi$ and $\hat{\alpha}$
\begin{equation*}
\hbar\langle  \zeta' |\hat{\alpha}|  \zeta'' \rangle=
(\overline{\psi}'-\overline{\psi}'')\langle  \zeta' |\psi|  \zeta'' \rangle\,.
\end{equation*}

Besides $p_{\mu}$, $Q$, $\overline{Q}$, the theory possesses another
conserved charge, namely, the energy, that corresponds to time shifts
and is described by the Hamiltonian operator
\begin{equation}
H=L=\frac{1}{2}\eta^{\mu\nu}p_{\mu}p_{\nu}\,.
\label{6-15}
\end{equation}
The conservation law $H=E=\const$ implies the following condition on the
stationary states
\begin{equation}
-\frac{\hbar^{2}}{2}\eta^{\mu\nu}
\frac{\partial }{\partial \eta^{\mu}}\frac{\partial }{\partial \eta^{\nu}}\Phi=E\Phi\,,
\label{6-16}
\end{equation}
that corresponds to the Schr\"odinger equation.

Thus, the realization space for the operators of a theory
under consideration is constructed in outline. The action of the
geometric coordinate operators $x^{\mu}$ on wave functions
can be obtained from the definition
$x^{\mu}=\eta^{\mu}-\dfrac{i}{2}\overline{\psi}\gamma^{\mu}\circ\psi$,
the last term causes the integral operator form of $x^{\mu}$.
The realization of the momentum $p_{\mu}$, conjugated to $x^{\mu}$
is a differential operator with respect to artificially introduced
variable $\eta^{\mu}$. Because of $[x^{\mu}, x^{\nu}]\ne 0$ now there is not
a simple connection between $x^{\mu}$ and $p_{\mu}$ like to
$p_{\mu}=-i\hbar\dfrac{\partial }{\partial x^{\mu}}$ (moreover, in quantum
theory the derivative with respect to $x^{\mu}$ is undefined at all).

Wave functions depend on the variables $\eta^{\mu}$ and $\psi$,
the operators $p_{\mu}$ and $Q$ are described by the derivations with respect to
to $\eta^{\mu}$ and $\psi$ (see (\ref{6-8})). Note, that in classical theory
the derivative $\left.\dfrac{\partial }{\partial \overline{\psi}}\right|_{\eta}$
does not coincide with the invariant operator $D$. From
\begin{equation*}
\begin{split}
\left.\frac{\partial }{\partial \overline{\psi}}\right|_{x}
\Phi(\eta(x, \psi, \overline{\psi}), \overline{\psi})
&=\left.\frac{\partial }{\partial \overline{\psi}}\right|_{\eta}\Phi+
\frac{i}{2}\gamma^{\mu}\psi\left.\frac{\partial }{\partial\eta^{\mu}}\right|_{\overline{\psi}}\Phi\\
&=\left.\frac{\partial }{\partial \overline{\psi}}\right|_{\eta}\Phi+
\frac{i}{2}\gamma^{\mu}\psi\left.\frac{\partial }{\partial x^{\mu}}\right|_{\overline{\psi}}\Phi\,,
\end{split}
\end{equation*}
we get
\begin{equation*}
\left.\frac{\partial }{\partial \overline{\psi}}\right|_{\eta}=
\left( \left.\frac{\partial }{\partial \overline{\psi}}\right|_{x}-
\frac{i}{2}\gamma^{\mu}\psi\left.\frac{\partial }{\partial x^{\mu}}\right|_{\overline{\psi}} \right)\Phi\,,
\end{equation*}
this combination differs from $D\Phi$ by a sign. Such an expression
is precisely a definition of the action of $Q$ on a superspace
(see \cite{wess}--\cite{ferrara}).

In the end of this section, it is worthwhile to remark that,
introducing the basis of the Hilbert space of states we use the new
variables $\eta^{\mu}$ and $\overline{\psi}$. The initial classical Lagrangian
can be written in terms of them as
\begin{equation*}
L=\frac{1}{2}\eta^{\mu\nu}p_{\mu}p_{\nu}=\frac{1}{2}\eta_{\mu\nu}v^{\mu}v^{\nu}\,,\quad
v^{\mu}=\partial\eta^{\mu}-i\partial\overline{\psi}\gamma^{\mu}\psi\,.
\end{equation*}
The corresponding momenta are
\begin{equation*}
\frac{\partial L}{\partial \partial\eta^{\mu}}=p_{\mu}=\eta_{\mu\nu}v^{\nu}\,,\quad
\frac{\partial L}{\partial \partial\overline{\psi}}=Q=-i\hat{p}\psi\,,\quad
\frac{\partial L}{\partial \partial\psi}=0\,,
\end{equation*}
i. e. the theory is constrained. In the present consideration we
have not introduced the phase space variables, therefore the constraint
on momenta $Q+i\hat{p}\psi=0$
transforms in a natural way to the into the definition of
the coordinate representation of the operator $\psi$.

In a similar manner one can construct the coordinate representation for the
total set $(\psi, \eta^{\mu}_{-})$. The basis, generated by these operators,
is connected with described above by means of a unitary transformation,
analogously to the connection between the coordinate and momentum
representations in usual quantum theory.


\section{Discussion}

In the preceding papers \cite{paper1}-\cite{paper3} and in the present
one we have considered the extension of Schwinger's quantization
procedure to the case of manifolds with a group structure including a
superspace. The presented approach may be viewed as effectual method of
study of some quantum models where the physical meaning of a theory
is closely connected with its geometric structure. Usually in these cases
the canonical quantization postulates are not self-consistent and its use
for constructing a theory requires some truthful but not sufficiently argued
assumptions. At the same time extended Schwinger's quantization procedure
based on symmetry properties of a theory expressed in terms of the
framework of Lie groups and algebras allow to determine the following:
\begin{description}
\item[(i)]
the quantum Lagrangian for a particle on the manifold with a group structure;
\item[(ii)]
quantum equations of motion for a particle and corresponding conservation laws;
\item[(iii)]
the algebra of commutation relations for operators describing a particle;
\item[(iv)]
the coordinate representation of quantum mechanics, i.~e. the form of the action
of operators on wave functions, the wave equation, etc.
\end{description}

In contradiction to the canonical quantization approach, Schwinger's
method in principle do not require the use of phase space variables.
Due to this reason it may be happen that in degenerate theories
the usual analyzes of constraints \cite{dirac} become unnecessary.

The results obtained in \cite{paper1}--\cite{paper3} and in the present paper
satisfy the correspondence principle and are connected in general
with the main results of works \cite{others1}-\cite{others2}, where the structure of a theory
has been obtained by the use of canonical quantization procedure supplemented
by additional special assumptions.

It seems that the conclusions derived from our series of works  may be
useful in investigations of quantum theory of a particle interacting with a
gravitational field and in solitonic models of elementary particles in the
framework of the collective-coordinate formalism.

\end{document}